\documentclass[10pt]{article}
\usepackage[utf8]{inputenc}
\usepackage{graphicx}
\usepackage[utf8]{inputenc}
\usepackage{amsmath}
\usepackage{amsfonts}
\usepackage{amssymb}
\usepackage{graphicx}

\usepackage{amsmath, physics, siunitx}

\usepackage{epstopdf}
\AppendGraphicsExtensions{.tif}

\usepackage{setspace}
\usepackage{authblk}
\usepackage{caption}
\usepackage{float}
\usepackage{cite}
\usepackage[left=2.0cm,right=2.0cm,top=2cm,bottom=2cm]{geometry}
\usepackage{color}
\usepackage{color,soul}
\usepackage{multicol}
\usepackage{subcaption}

\title{\textbf{Turbulence Impacted Beam Statistics and Image Topology with Lorentz Dipole Oscillation}}

\author[1,*]{Shouvik Sadhukhan}

\author[2]{C. S. Narayanamurthy}

\affil[1, 2]{\small{Applied and Adaptive Optics Laboratory, Department of Physics, Indian Institute of Space Science and Technology (IIST), P.O: Valiamala, Trivandrum - 695547, State: Kerala; India}}
\affil[1]{\small{Email: shouvikphysics1996@gmail.com}}
\affil[2]{\small{Email: naamu.s@gmail.com}}
\affil[*]{\small{Corresponding Author Email: shouvikphysics1996@gmail.com}}

\begin{document}
\maketitle

\begin{abstract}
This work presents a rigorous statistical and geometric framework for analyzing turbulence-impacted beam propagation and image topology with results obtained using a PMMA slab. The approach models beam intensity distributions as n-dimensional data set represented through Gaussian Mixture Models (GMMs), embedding them into the manifold of Symmetric Positive Definite (SPD) matrices. By employing information geometric tools, geodesic distances, and affine-invariant Riemannian metrics, we establish a principled methodology for quantifying similarity and dissimilarity among beam images. Experimental results demonstrate topological distance trends, distance statistics, and correlation measures for different turbulence scenarios, including polarized and unpolarized cases. Histograms of distance statistics reveal stable statistical features, with correlation coefficients highlighting the turbulence-induced variability in PMMA-based beam propagation. The framework not only provides a systematic foundation for analyzing optical beam statistics under turbulence but also opens avenues for advanced applications such as deep learning-based feature reduction, image compression, and secure free-space optical (FSO) communication. Future directions include refining the GMM-EM based distance measures, comparative scatter analysis, and developing robust statistical tools for turbulence imaging. Overall, this study bridges theoretical modeling, experimental validation, and potential technological applications in adaptive and applied optics.\\

\end{abstract}

\section{Introduction}
Free-space optical communication systems are highly vulnerable to atmospheric turbulence, which introduces random fluctuations in intensity, phase, and polarization. These distortions manifest as beam wander, scintillation, and coherence loss, severely impacting link reliability. To address this, the present study develops a unified theoretical and experimental framework that exploits collective dipole oscillations within dense dielectric media. By harnessing the synchronized response of coupled dipoles, the proposed approach effectively compensates turbulence-induced degradation, offering a promising pathway toward robust optical communication in realistic environments.\cite{6,7,8,9,10,11,12,13,14,15,16,17,18,19,20}

This body of work spans theoretical, computational, and experimental advances that deepen our understanding of optical forces, electromagnetic dipoles, and wave–matter interactions. On the mathematical side, Vishwakarma and Moosath introduced distance measures for Gaussian mixtures, offering new tools for statistical signal analysis \cite{1}. Korotkova and collaborators systematically explored intensity fluctuations, coherence, and polarization statistics of random beams in turbulent media, establishing how environmental randomness governs beam evolution \cite{2,3,4,5,6,7,8,9,10}. Parallel efforts in Jones and Mueller matrix calculus extended the formalism for characterizing polarized beams \cite{11,12,13,14,15}. Experimental frontiers include Abbasirad et al.’s nanoscale studies of dipole emission near metasurfaces \cite{16}, while Scheel and Buhmann \cite{17} provided a macroscopic quantum electrodynamics framework for complex media. Classic works by Levine, Schwinger, and Sipe laid the theoretical foundation for diffraction, scattering, and dielectric resonances \cite{18–30}. A second cluster of contributions addresses dipole–dipole interactions and optical forces. Poddubny et al. modeled Purcell enhancement in hyperbolic metamaterials \cite{20}, while Martin, Piller, and Paulus developed accurate scattering and Green’s tensor techniques \cite{21,22}. Extensions of the Lorentz oscillator with nonlinear and damping effects \cite{23,29}, alongside studies of light-induced dipole forces in gases and structured fields \cite{33–40}, highlight optomechanical applications. Additional works on discrete dipole approximations, quadrupole effects, and nonlinear metasurfaces \cite{25,26,30,45} collectively establish a multidisciplinary framework guiding current and future optical research.\par

Research on turbulence-impacted beam statistics and image topology draws upon extensive developments in statistical inference, information geometry, and optical physics. Deep learning frameworks by Goodfellow et al.\cite{66} and Bengio et al.\cite{71} provide methods for feature extraction and dimensionality reduction, supporting complex turbulence data analysis. Probabilistic modeling has been strengthened through divergence measures, with Pardo\cite{67} and the seminal work of Kullback and Leibler\cite{76}, while Amari and Nagaoka\cite{68} expanded these concepts via Riemannian geometry for comparing statistical manifolds. Foundational texts on matrix computations\cite{69} and smooth manifolds\cite{72} further provide the mathematical basis. Within Gaussian Mixture Model (GMM) approaches, Durrieu et al.\cite{74} and Hershey and Olsen\cite{85} developed approximations for KL divergence, while Kampa et al.\cite{77} and Calvo and Oller\cite{78} proposed alternative metrics. More recent studies introduced geometry-aware similarity measures\cite{79,80}, earth mover’s distance formulations\cite{86}, and GMM-based classification techniques\cite{75,87}. Applications to image similarity include KL-based retrieval and texture statistics by Goldberger et al.\cite{84} and Kwitt and Uhl\cite{82}, and fractal-invariant methods by Xu et al.\cite{83}. Complementary advances in optical physics, such as Genco et al.’s femtosecond switching of strong light–matter coupling\cite{89} and Peters and Rodriguez’s precision nonlinear sensing\cite{90}, emphasize how statistical-geometric tools extend beyond turbulence, offering broad relevance to modern optics and photonics.\par

Optical beam propagation through random media, such as atmospheric turbulence or disordered dielectric materials, introduces significant challenges for both theory and applications by causing scintillation, beam wander, wavefront distortion, and topological changes that degrade free-space optical communication, imaging, and adaptive optics performance. Conventional statistical tools—like scintillation index, correlation coefficients, and kernel density estimation—capture limited aspects of these distortions but lack a unified geometric or topological perspective. To address this, we develop a rigorous framework that combines statistical modeling with information geometry for analyzing turbulence-affected beams. In this approach, experimentally obtained intensity distributions are treated as n-dimensional data set and modeled using Gaussian Mixture Models (GMMs), which are then embedded into the manifold of Symmetric Positive Definite (SPD) matrices. This embedding enables the use of affine-invariant Riemannian metrics and geodesic distances, providing a principled method to quantify similarities and differences in beam topology across turbulence conditions. Experimentally, turbulence is studied using polymethyl methacrylate (PMMA) slabs, with Gaussian laser beams recorded under turbulence-free, raw turbulence, and turbulence-compensated scenarios using one or two PMMA rods. The SPD-based analysis reveals topological distance trends and correlation measures, showing how compensation strategies mitigate turbulence effects. Beyond immediate validation, this framework has potential for machine learning, secure optical encryption, data compression, and robust turbulence imaging.\par

When an electromagnetic wave passes through a transparent or semi-transparent medium, its electric field perturbs the electron clouds of the constituent molecules, breaking their stationary symmetry and inducing oscillating dipoles. These dipoles do not behave independently; instead, they interact and exchange energy, giving rise to coupled oscillations. The resulting coupling introduces non-diagonal terms into the governing equations, so the dynamics cannot be expressed in simple Cartesian coordinates. To address this, diagonalization is applied, redefining the system in terms of new orthonormal modes. This transformation uncovers collective oscillations, where dipoles synchronize and exhibit coherent dynamics, in contrast to the initial random, uncorrelated motion. The diagonalized modes thus capture ordered, emergent behavior arising from microscopic interactions. By highlighting the transition from chaotic individual oscillations to stabilized collective modes, this framework provides a clear understanding of how local dipole coupling governs the macroscopic electromagnetic response of the medium.
\cite{56,57,58,59,60,61,62,63,64,65,66,67,68,69,70,71,72,73,74,75} \par

The present work demonstrated as in section \ref{2} detail theoretical discussion of the work have been given. The statistical background has been discussed in section \ref{3}. The section \ref{4} contains experimental details where the results analysis have been added in section \ref{5}. Finally, the paper is concluded into section \ref{6}.

\section{Theoretical Background}
The fundamental theoretical work begings with the dynamical equation of the Lorentz Oscillation model under external field. As in the present framework, we are propagating a light beam through a medium. After the entrance of that beam, it generates dipoles inside the material. The dipole-dipole coupling causes the intra-medium electromagnetic field propagation. On the other hand, the dipole-dipole coupling influences the resultant oscillation modes of the dipoles when a randomly polarized light generates randomly distributed dipoles inside the medium. Thus, the final differential equation of Generalized Anharmonic Dipole Oscillator with Dipole-Dipole Coupling can be written as follows with dipole-dipole coupling coefficient as Dyadic Greens Function $\mathbf{G}(\mathbf{r}_i, \mathbf{r}_j)$.

\begin{equation}
    m \ddot{\mathbf{r}}_i + m \gamma \dot{\mathbf{r}}_i + m \omega_0^2 \mathbf{r}_i + \beta_i |\mathbf{r}_i| \mathbf{r}_i + \alpha_i |\mathbf{r}_i|^2 \mathbf{r}_i = -e \mathbf{E}_{\text{ext}}(\mathbf{r}_i, t) + e^2 \mu_0 \omega^2 \sum_{j \ne i} \mathbf{G}(\mathbf{r}_i, \mathbf{r}_j) \cdot \mathbf{r}_j(t)
\end{equation}

Here $\vec{\textbf{r}}$ is the relative position vector with respect to the nucleus, $m\rightarrow$ effective electron mass where, $\frac{1}{m}=\frac{1}{\hbar^2}\frac{\partial^2 E(\kappa)}{\partial\kappa^2}$ with $E(\kappa)$ is the electron dispersion relation which is generally defined by the curvature of the electronic band structure, $\omega_0 \rightarrow$ natural (resonant) frequency of bound electron oscillation and $\gamma\rightarrow$ damping constant (collisions, radiation loss). The term which is contained with natural frequency represents the restoring effect on the external field which causes the perturbation on the electron cloud to induce dipole moment. $\beta_i\rightarrow$ second-order nonlinearity coefficient and $\alpha_i\rightarrow$ third-order nonlinearity coefficient. Whenever an electromagnetic field enters into a transparent or semi-transparent medium, the electric field of the EM wave interacts with the molecules of the medium. The interaction between the electric field and the material/medium molecules produces an electric field perturbation on the electron clouds. The presence of perturbations breaks the stationary symmetries of the molecules and the corresponding electron clouds. Thus, the breaking of symmetries generates dipoles inside the molecules. The oscillatory amplitude variation of the external field provides dipole oscillations inside the molecules of the propagating medium. The molecules inside the material are inter-connected and hence, they can exchange energy. Thus, coupled dipole oscillation model comes into scenario. $\overline{\overline{G}}$ generalizes the scalar Green’s function to vector fields, ensuring transversality and coupling between components. The solution of the green's differential equation $\left[ \nabla \times \nabla \times - k^2 \mathbf{I} \right] 
\overline{\overline{G}}(\mathbf{r},\mathbf{r}') 
= \mathbf{I} \delta(\mathbf{r}-\mathbf{r}')$ can be derived as follows which basically produces the expanded scalar green's function $g(\mathbf{r},\mathbf{r}') = \frac{e^{ik|\mathbf{r}-\mathbf{r}'|}}{4\pi |\mathbf{r}-\mathbf{r}'|}$.

\begin{equation}
    \overline{\overline{G}}(\mathbf{r},\mathbf{r}') 
= \left( \mathbf{I} + \frac{1}{k^2} \nabla \nabla \right) \frac{e^{ik|\mathbf{r}-\mathbf{r}'|}}{4\pi |\mathbf{r}-\mathbf{r}'|}
\end{equation}

The gradient of the electric field with non-uniform amplitude produces an additional impact on the coupling between the nearest neighbour dipoles. The presence of non-zero gradients and higher-orders derivatives of the electric field produces non-uniform dipole moments on spatially distributed dipoles. These non-uniformities modify the conventional dipole-dipole coupling by introducing gradient-based repelling forces. Hence, the gradient force contribution should be considered into the generalized Lorentz force dynamics. In the present context of the formalism for the Generalized Anharmonic Dipole Oscillator with Dipole-Dipole Coupling we conventionally found two forces that are modulating the localized dipole moments. Those forces are as follows.
\begin{itemize}
    \item $\mathbf F_{Ext}=-e \mathbf{E}_{\text{ext}}(\mathbf{r}_i, t)$ which is the externally applied field due to propagation of light fields.
    \item $\mathbf F_{Coupling}= e^2 \mu_0 \omega^2 \sum_{j \ne i} \mathbf{G}(\mathbf{r}_i, \mathbf{r}_j) \cdot \mathbf{r}_j(t)$ which is the dipole-dipole coupled forces that are applied from surrounding localized dipoles on the specific dipole.
\end{itemize}
Now, if the external force field i.e., the propagating field spatial distribution is non-uniform or variable, the external force field can be expanded into corresponding Taylor's expansion. The higher order derivatives of the spatially varying external field will contribute additional force on coupling. That forces act as Gradient Coupling Force. Hence, total force on a point $r_i$ can be written as follows ($HO=$ Higher orders).

\begin{equation}
    \mathbf F_{TotExt}=-e \mathbf{E}_{\text{ext}}(\mathbf{r}_{i}, t)-\sum_{j \ne i}\left [(e(\vec{\mathbf r}_i -  \vec{\mathbf  r}_j)\cdot\nabla)\mathbf{E}_{\text{ext}}(\mathbf{r}_{j}, t)+ HO\right]=\mathbf F_{Ext}+\mathbf F_{HO}
\end{equation}

Thus, in the present context, the total coupling forces can be written as a summation of Gradient Coupling Force and Dipole-Dipole Coupling Forces as follows.

\begin{equation}
    \mathbf F_{TotCoupling} = \mathbf F_{Coupling} + \mathbf F_{HO} = e^2 \mu_0 \omega^2 \sum_{j \ne i} \mathbf{G}(\mathbf{r}_i, \mathbf{r}_j) \cdot \mathbf{r}_j(t) +\sum_{j \ne i}\left ((e(\vec{\mathbf r}_j -  \vec{\mathbf  r}_i)\cdot\nabla)\mathbf{E}_{\text{ext}}(\mathbf{r}_{j}, t)+ HO\right)
\end{equation}

This total coupling force will insist the dipole system to be synchronized to have oscillation with common diagonalized modes. Thus, after saturation of the synchronization, the sudden change in the 2D spatial field distribution can't change the synchronized oscillation modes. Hence, the centroid shift will slow down. The final representation of the Lorentz dynamics including gradient forces should be presented as follows.

\begin{eqnarray}\label{15}
    && m \ddot{\mathbf{r}}_i + m \gamma \dot{\mathbf{r}}_i + m \omega_0^2 \mathbf{r}_i + \beta_i |\mathbf{r}_i| \mathbf{r}_i + \alpha_i |\mathbf{r}_i|^2 \mathbf{r}_i \nonumber\\ &&= -e \mathbf{E}_{\text{ext}}(\mathbf{r}_i, t) + e^2 \mu_0 \omega^2 \sum_{j \ne i} \mathbf{G}(\mathbf{r}_i, \mathbf{r}_j) \cdot \mathbf{r}_j(t) +\sum_{j \ne i}\left ((e(\vec{\mathbf r}_j -  \vec{\mathbf  r}_i)\cdot\nabla)\mathbf{E}_{\text{ext}}(\mathbf{r}_{j}, t)+ HO\right)
\end{eqnarray}

The presence of coupling in the dipole dynamical system contains constraints which modifies the degrees of freedom. This system can't produce independent dynamical equations and hence, diagonalization is needed to apply on the coupled dipole dynamical equations. We define the collective displacement vector as $R(t)=\big[r_1(t),r_2(t),\dots,r_N(t)\big]^T$ with the dipole–dipole interaction matrix $C_{ij} = 
\begin{cases}
-\,k_0^2\mu_0\omega^2 G(r_i,r_j), & i\neq j, \\
0, & i=j.
\end{cases}$. Now following the paper \cite{91} we can find the diagonalized equation as follows.

\begin{equation}
\Big[-\omega^2I+i\gamma\omega I+\Lambda+\tfrac{1}{m}\widetilde B\Big]Q_\omega
= -\frac{e}{m}U^{-1}E_\omega+\frac{1}{m}U^{-1}F_{{\rm grad},\omega},
\label{eq:modal}
\end{equation}

where $B^{(1)}_2=\mathrm{diag}(\beta_i\sqrt{2}|r_{\omega,i}|)$ and $B^{(1)}_3=\mathrm{diag}(3\alpha_i|r_{\omega,i}|^2)$. Diagonalize the stiffness matrix $K_{\rm eff}=U\Lambda U^{-1},\qquad \Lambda=\mathrm{diag}(\Omega_1^2,\dots,\Omega_{3N}^2)$. Thus, the radiated output field is expressed as

\begin{equation}
E_{\rm out}(r)=k_0^2\varepsilon_0\int\!\!\int 
G(r,r')\,\chi(r',r'';\omega,|E_\omega|)\,E_\omega(r'')\,d^3r''\,d^3r',
\end{equation}

where the effective susceptibility kernel contains the nonlinear and modal corrections. In modal form, inserting Eq.\eqref{eq:modalzero}, we obtain

\begin{equation}
E_{\rm out}(r)=
k_0^2\varepsilon_0\sum_n \phi_n(r)\,
\frac{Ne^2}{\varepsilon_0 m}\,
\frac{\langle\phi_n|E_\omega\rangle-\tfrac{1}{e}\langle\phi_n|F_{{\rm grad},\omega}\rangle}
{\Omega_n^2-\omega^2-i\gamma\omega}
\;\otimes\;
\Big(\int\!\!\int \phi_n(r')\otimes\phi_n^*(r'')\,G(r,r')\,E_\omega(r'')\,d^3r''\,d^3r'\Big).
\label{eq:outputfinal}
\end{equation}

In the present context, the presence of dynamic turbulence impact on the electric field spatial distribution of the propagating optical field includes the time dependency of this perturbation force. Hence, we must have $\delta \mathbf F_{Pert}\rightarrow\delta\mathbf F_{Pert}(t)$. The perturbation forces can be summed as follows when the field distribution changes after synchronization.

\begin{eqnarray}
    &&\mathbf{\delta F}_{Pert} = \mathbf F_{Tot} - \mathbf F'_{Tot} = \mathbf F_{Ext} + \mathbf F_{Coupling} + \mathbf F_{HO} - \mathbf F'_{Ext} - \mathbf F'_{Coupling} - \mathbf F'_{HO} - \mathbf F_{Lorentz} \nonumber\\&&= -e \mathbf{E}_{\text{ext}}(\mathbf{r}_i, t) + e^2 \mu_0 \omega^2 \sum_{j \ne i} \mathbf{G}(\mathbf{r}_i, \mathbf{r}_j) \cdot \mathbf{r}_j(t) +\sum_{j \ne i}\left ((e(\vec{\mathbf r}_j -  \vec{\mathbf  r}_i)\cdot\nabla)\mathbf{E}_{\text{ext}}(\mathbf{r}_{j}, t)+ HO\right)\nonumber\\&& +e \mathbf{E'}_{\text{ext}}(\mathbf{r}_i, t) - e^2 \mu_0 \omega^2 \sum_{j \ne i} \mathbf{G'}(\mathbf{r}_i, \mathbf{r}_j) \cdot \mathbf{r}_j(t) -\sum_{j \ne i}\left ((e(\vec{\mathbf r}_j -  \vec{\mathbf  r}_i)\cdot\nabla)\mathbf{E'}_{\text{ext}}(\mathbf{r}_{j}, t)+ HO\right) \nonumber\\&&- (\mathbf p_i\cdot\nabla)\mathbf E'_{Ext}(\mathbf r_i, t) - \dot{\mathbf p}_i\times \mathbf B' (\mathbf r_i,t) - HOMP'
\end{eqnarray}

After such perturbation, the dynamical equation must look like the following equation.

\begin{eqnarray}
    && m \ddot{\mathbf{r}}_i + m \gamma \dot{\mathbf{r}}_i + m \omega_0^2 \mathbf{r}_i + \beta_i |\mathbf{r}_i| \mathbf{r}_i + \alpha_i |\mathbf{r}_i|^2 \mathbf{r}_i \nonumber\\ &&= -e \mathbf{E}_{\text{ext}}(\mathbf{r}_i, t) + e^2 \mu_0 \omega^2 \sum_{j \ne i} \mathbf{G}(\mathbf{r}_i, \mathbf{r}_j) \cdot \mathbf{r}_j(t) +\sum_{j \ne i}\left ((e(\vec{\mathbf r}_j -  \vec{\mathbf  r}_i)\cdot\nabla)\mathbf{E}_{\text{ext}}(\mathbf{r}_{j}, t)+ HO\right) +  \delta \mathbf{F}_{Pert} (t)
\end{eqnarray}

We can find the following conditions on the perturbed force depending upon corresponding magnitudes with $\delta \mathbf{F}_{Pert} =  \mathbf F'_{Inertia} - \mathbf F_{Inertia}$.
\begin{itemize}
    \item $\delta\mathbf F_{Pert}\rightarrow 0$: The output field distribution can be caused by the saturated synchronized dipole moment distribution. Hence, the turbulence impact can be compensated fully with presence of medium dipole-dipole coupling energy transitions.
    \item $\delta\mathbf F_{Pert}\rightarrow $Small but $\neq 0$: The dynamic nature of turbulence changes the synchronization with induction of perturbed inertia force. The perturbation is small, hence, output field turbulence impact can be found compensated.
    \item $\delta\mathbf F_{Pert}>0$: In such case, the output field distribution will be dependent upon the frequency of change of perturbation. If turbulence is strong, i.e., change of perturbation is rapid, the medium dipole coupled system can't find time to be synchronized. Thus, for strong turbulence, we can find un-compensated turbulence impacted output field. If the turbulence is weak, output field can be found compensated from turbulence.
\end{itemize}

\section{Statistical Background}\label{3}
In this section we provide a detailed theoretical framework for analyzing higher-order statistics of two-dimensional intensity images using cumulants. The method begins with the interpretation of intensity values as a probability measure, proceeds through mean and covariance estimation, performs Cholesky whitening for standardization, and finally computes higher-order cumulants and employs a Gram--Charlier expansion to model deviations from Gaussianity. Given a discrete nonnegative intensity map $I_{ij}$ on pixels $(x_j,y_i)$, define 
weighted values $W_{ij}$ and normalization
\begin{equation}
S=\sum_{i,j}W_{ij}, \qquad p_{ij}=\frac{W_{ij}}{S}, \qquad \sum_{i,j}p_{ij}=1,
\end{equation}
so that expectations of functions $f(X,Y)$ follow
\begin{equation}
\mathbb{E}[f(X,Y)]=\sum_{i,j}p_{ij} f(x_j,y_i).
\end{equation}

The centroid and covariance are
\begin{equation}
\mu_x=\mathbb{E}[X], \quad \mu_y=\mathbb{E}[Y], \qquad
\Sigma=\begin{pmatrix}\sigma_{xx} & \sigma_{xy}\\ \sigma_{xy} & \sigma_{yy}\end{pmatrix}.
\end{equation}
With Cholesky factorization $\Sigma=LL^\top$, whitened coordinates
\begin{equation}
z=L^{-1}\begin{pmatrix}x-\mu_x \\ y-\mu_y\end{pmatrix}, \qquad 
\mathbb{E}[z]=0, \quad \mathrm{Cov}(z)=I,
\end{equation}
remove correlations and scale anisotropy.

Standardized moments from whitened samples $(z_1,z_2)$ are
\begin{equation}
m_{pq}=\sum_k w_k z_{1k}^p z_{2k}^q, \qquad 
m_{20}=m_{02}=1,\; m_{11}=0,
\end{equation}
with higher orders $m_{30},m_{21},\ldots,m_{04}$.  
Third-order cumulants equal moments:
\begin{equation}
k_{30}=m_{30},\; k_{21}=m_{21},\; k_{12}=m_{12},\; k_{03}=m_{03},
\end{equation}
and fourth-order reduce to
\begin{equation}
k_{40}=m_{40}-3,\quad k_{04}=m_{04}-3,\quad 
k_{31}=m_{31},\quad k_{13}=m_{13},\quad k_{22}=m_{22}-1.
\end{equation}
Global skewness and kurtosis norms are
\begin{align}
|\text{skew}|_3 &= \sqrt{k_{30}^2+k_{21}^2+k_{12}^2+k_{03}^2},\\
|\text{kurt}|_4 &= \sqrt{k_{40}^2+k_{31}^2+k_{22}^2+k_{13}^2+k_{04}^2}.
\end{align}

In whitened coordinates, the Gaussian density is
\begin{equation}
\phi(z)=(2\pi)^{-1}\exp(-|z|^2/2),
\end{equation}
with Hermite polynomials $H_\alpha(z)$ (e.g.\ $H_{30}=z_1^3-3z_1$, $H_{22}=z_1^2z_2^2-z_1^2-z_2^2+1$).  
The Gram--Charlier expansion is
\begin{align}
p(r) &\approx \phi(r)\Big[1+\tfrac{1}{6}(k_{30}H_{30}+3k_{21}H_{21}+3k_{12}H_{12}+k_{03}H_{03}) \nonumber\\
&\quad+\tfrac{1}{24}(k_{40}H_{40}+4k_{31}H_{31}+6k_{22}H_{22}+4k_{13}H_{13}+k_{04}H_{04})\Big].
\end{align}

To match the expansion to measured image intensity, we include a scaling and offset:

\begin{equation}
I_{\rm fit}(r)=a\,p(r)+b,
\end{equation}

With $a,b$ fitted by least squares to capture global amplitude and background, the method extracts higher-order statistics (skewness, kurtosis excess) from 2D intensity images. The pipeline is: (i) normalize intensity to define a probability distribution; (ii) compute mean $\mu$ and covariance $\Sigma$; (iii) apply Cholesky whitening to obtain standardized coordinates $z$; (iv) evaluate standardized moments and convert to cumulants; (v) quantify skewness and kurtosis norms as measures of non-Gaussianity; (vi) incorporate cumulants into a Gram--Charlier expansion, fitted with scale and offset to model intensity beyond the Gaussian approximation. In the present theoretical frame work we have fitted our 2D optical beam images using 2D bi-variate Gaussian function with skewness and Kurtosis as noise distortions. After dynamic turbulence impact, the gaussian beam spatial distribution get expanded. The centroid starts to move. Hence, for every frame of measurements, the images become unbound in structures. Thus, fitted function bounded volume can become an important parameter to identify the structural changes in each frame in presence of dynamic turbulence impact. The volume can be formulized as follows.

\begin{equation}
    V_{Frame}=\int_{\mathbb R^2} I_{\rm }(\mathbf r,t)\,d^2\mathbf r=\int_{\mathbb R^2} \frac{I_{\rm Pixel}(\mathbf r,t)}{I_{\rm Max}(\mathbf r,t)}\,d^2\mathbf r
\end{equation}

In the information geometric framework, n-dimensional data sets are modelled as samples from a Gaussian Mixture Model (GMM), allowing their representation on a statistical manifold. For a n-dimensional data set $X = \{x_i \in \mathbb{R}^n\}_{i=1}^N$, the underlying probability density function is given by

\begin{equation}
    p(x; \theta) = \sum_{k=1}^K \pi_k \mathcal{N}(x \mid \mu_k, \Sigma_k), \quad \sum_{k=1}^K \pi_k = 1
\end{equation}

where $\theta = \{(\pi_k, \mu_k, \Sigma_k)\}_{k=1}^K$ parameterizes the manifold $\mathcal{M}$. To measure similarities between such models, an embedding $f: \mathcal{M} \to SPD_{K(n+1)}(\mathbb{R})$ is defined, where $SPD_m(\mathbb{R})$ denotes the manifold of $m \times m$ symmetric positive definite matrices. The embedding is

\begin{equation}
    f(\theta) = \begin{pmatrix}
A & X \\
X^T & B
\end{pmatrix}
\end{equation}

with

\begin{equation}
    A = diag(\Sigma_1 + \pi_1 \mu_1 \mu_1^T, \dots, \Sigma_K + \pi_K \mu_K \mu_K^T) \in \mathbb{R}^{Kn \times Kn}
\end{equation}

\begin{equation}
    X = diag(\pi_1 \mu_1, \dots, \pi_K \mu_K) \in \mathbb{R}^{Kn \times K}
\end{equation}
\begin{equation}
    B = diag(\pi_1, \dots, \pi_K) \in \mathbb{R}^{K \times K}
\end{equation}
This block matrix is symmetric positive definite, embedding the GMM into the SPD manifold. The geometry of $SPD_n(\mathbb{R})$ is equipped with a Riemannian metric
\begin{equation}
    d s^2 = \frac{1}{2} \Tr\left( (S^{-1} d S)^2 \right)
\end{equation}
for $S \in SPD_n(\mathbb{R})$ and $d S$ an infinitesimal perturbation in the tangent space. The affine-invariant metric on $SPD_n(\mathbb{R})$ is defined for $A \in SPD_n(\mathbb{R})$ and $X, Y \in T_A SPD_n(\mathbb{R})$ as
\begin{equation}
    \rho_A(X, Y) = \Tr(A^{-1} X A^{-1} Y)
\end{equation}
This metric is invariant under transformations $A \mapsto G^T A G$ for $G \in GL_n(\mathbb{R})$. The geodesic path connecting $A, B \in SPD_n(\mathbb{R})$ is
\begin{equation}
    \gamma(t) = A^{1/2} (A^{-1/2} B A^{-1/2})^t A^{1/2}, \quad t \in [0, 1]
\end{equation}

The corresponding geodesic distance, known as the Affine-Invariant Riemannian Metric (AIRM) distance, is
\begin{equation}
    d(A, B) = \left\| \log(A^{-1/2} B A^{-1/2}) \right\|_F
\end{equation}
where $\log$ denotes the matrix logarithm and $\|\cdot\|_F$ is the Frobenius norm. This distance quantifies the dissimilarity between embedded n-dimensional data set on the SPD manifold, facilitating analysis of geometric and statistical properties in turbulence-impacted optical beams. By Theorem~3 from reference \cite{1}, the embedding $f:\mathcal{M}\to \mathrm{SPD}_{K(n+1)}(\mathbb{R})$ makes
$(\mathcal{M},f^{*}\rho)$ isometric to its image $(f(\mathcal{M}),\rho|_{{f(\mathcal{M})}})$, so the pullback of the affine invariant Riemannian Metric endows the GMM manifold with a Riemannian metric. By Theorem~4 from reference \cite{1}, the affine invariant Riemannian distance on an ambient space is a \emph{lower bound} for the the pullback distance on the manifold of GMMs. This enables one to use the affine invariant Riemannian distance on the ambient space for the similarity measure for GMMs. In analyzing the statistical properties of turbulence-impacted optical beams, Pearson correlation provides a measure of similarity between probability density functions derived from beam intensity distributions. For two probability density functions $f(x)$ and $g(x)$ sampled over bins $x_i$ ($i = 1, \dots, N$), the Pearson correlation coefficient is defined as
\[
\rho_{f,g} = \frac{\sum_{i=1}^N \left( f(x_i) - \bar{f} \right) \left( g(x_i) - \bar{g} \right)}{\sqrt{\sum_{i=1}^N \left( f(x_i) - \bar{f} \right)^2} \sqrt{\sum_{i=1}^N \left( g(x_i) - \bar{g} \right)^2}},
\]\begin{equation}
    \rho_{f,g} = \frac{\sum_{i=1}^N \left( f(x_i) - \bar{f} \right) \left( g(x_i) - \bar{g} \right)}{\sqrt{\sum_{i=1}^N \left( f(x_i) - \bar{f} \right)^2} \sqrt{\sum_{i=1}^N \left( g(x_i) - \bar{g} \right)^2}}
\end{equation}
where $\bar{f} = \frac{1}{N} \sum_{i=1}^N f(x_i)$ and $\bar{g} = \frac{1}{N} \sum_{i=1}^N g(x_i)$ are the means of $f$ and $g$, respectively. The coefficient $\rho_{f,g}$ ranges from $-1$ to $1$. The value of $\rho_{f,g}$ offers insights into the relationship between $f$ and $g$:
- A positive correlation ($\rho > 0$) indicates that $f$ and $g$ vary in the same direction, with peaks and troughs aligning, suggesting greater similarity as $\rho$ increases.
- A negative correlation ($\rho < 0$) implies variation in opposite directions, where peaks of one coincide with troughs of the other, indicating structural dissimilarity.
- A correlation near zero ($\rho \approx 0$) suggests no clear linear relation, with fluctuations appearing independent. To estimate the underlying probability density functions from data points, Kernel Density Estimation (KDE) is employed. Given $n$ samples $\{x_1, x_2, \dots, x_n\}$ from an unknown distribution, the KDE estimator at point $x$ is
\[
\hat{f}_h(x) = \frac{1}{n h} \sum_{i=1}^n K\left( \frac{x - x_i}{h} \right),
\]\begin{equation}
    \hat{f}_h(x) = \frac{1}{n h} \sum_{i=1}^n K\left( \frac{x - x_i}{h} \right)
\end{equation}
where $K(\cdot)$ is the kernel function (e.g., Gaussian or Epanechnikov) and $h > 0$ is the bandwidth, controlling the smoothing. A small $h$ leads to high variance and overfitting, while a large $h$ results in high bias and oversmoothing. KDE approximates the true underlying probability density function, facilitating the computation of correlations in turbulence-affected beam statistics.

\section{Experimental Varifications}\label{4}
\begin{figure}[h]
\centering
\begin{minipage}[b]{0.75\textwidth}
    \includegraphics[width=\textwidth]{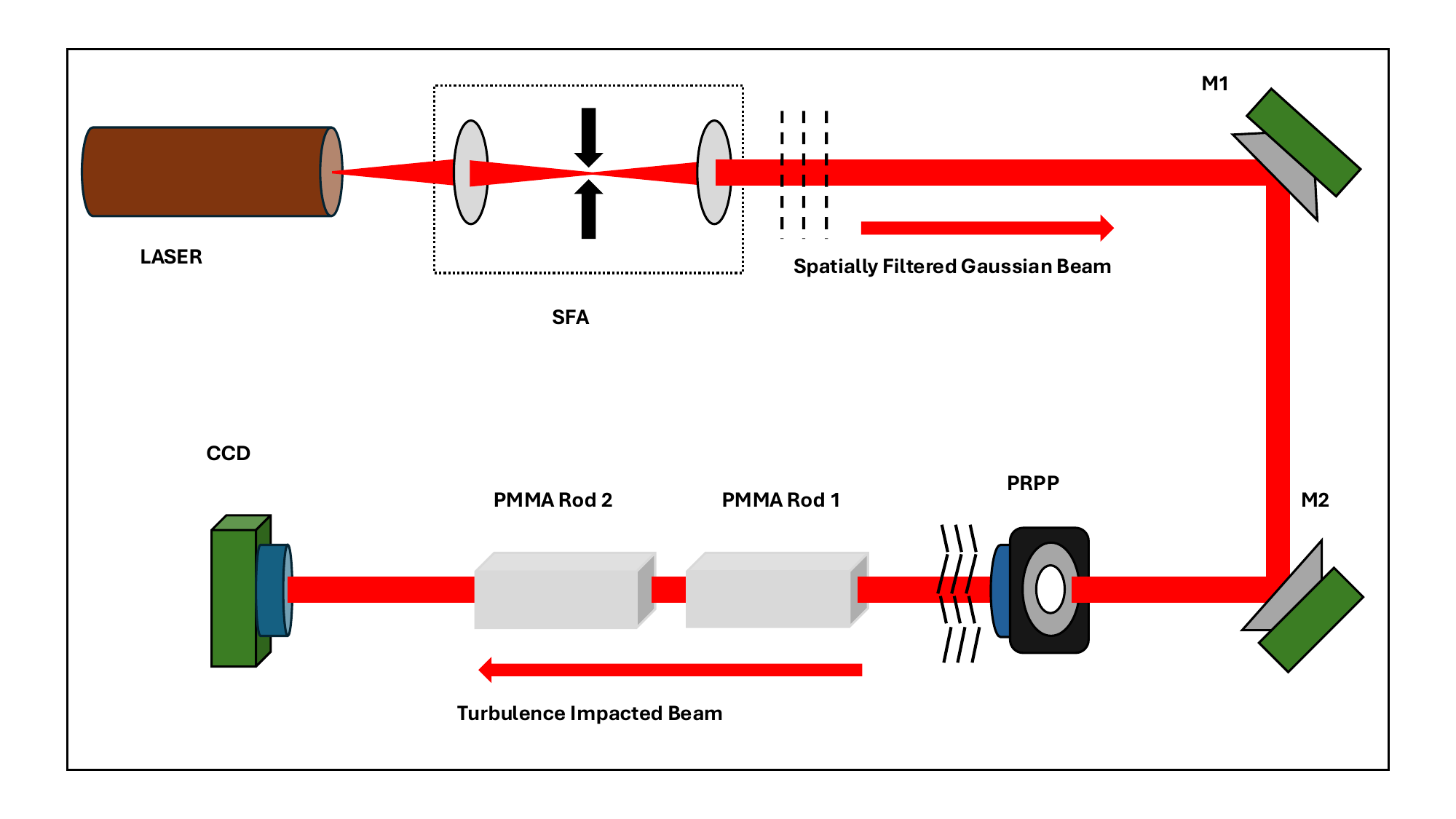}
    \caption{Experimental procedure with 2 PMMA Rod}
    \label{P0}
\end{minipage}
\end{figure}

The experimental setup (Fig.\ref{P0}) begins with a continuous-wave laser passed through a spatial filter assembly (SFA) to produce a clean Gaussian beam by suppressing higher-order distortions. Mirrors M1 and M2 guide and align the beam, which then passes through a programmable rotating phase plate (PRPP) introducing controlled turbulence via dynamic random phase modulation. The turbulence-affected beam propagates through one or two polymethyl methacrylate (PMMA) rods, enabling study of cumulative light–matter interaction effects. Finally, a CCD camera records the transmitted field, capturing both turbulence-impacted and compensated intensity distributions for subsequent analysis. (Figure of experimental set-up is similar as reference \cite{91})

\begin{figure}[h]
\centering
\begin{minipage}[b]{0.45\textwidth}
    \includegraphics[width=\textwidth]{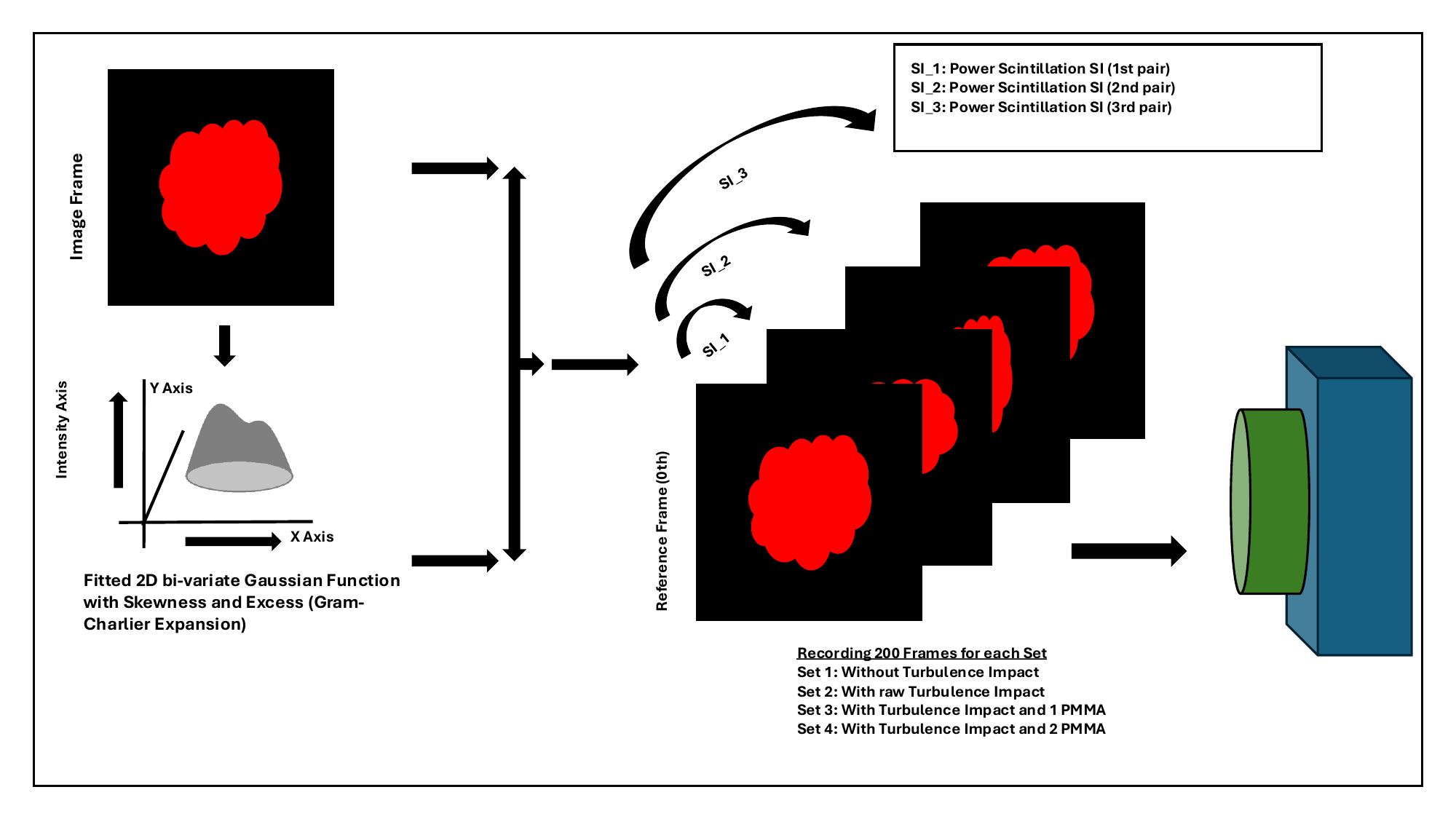}
    \subcaption{Block diagram Data Analysis Scheme I}
    \label{P1a}
\end{minipage}\hfill
\begin{minipage}[b]{0.50\textwidth}
    \includegraphics[width=\textwidth]{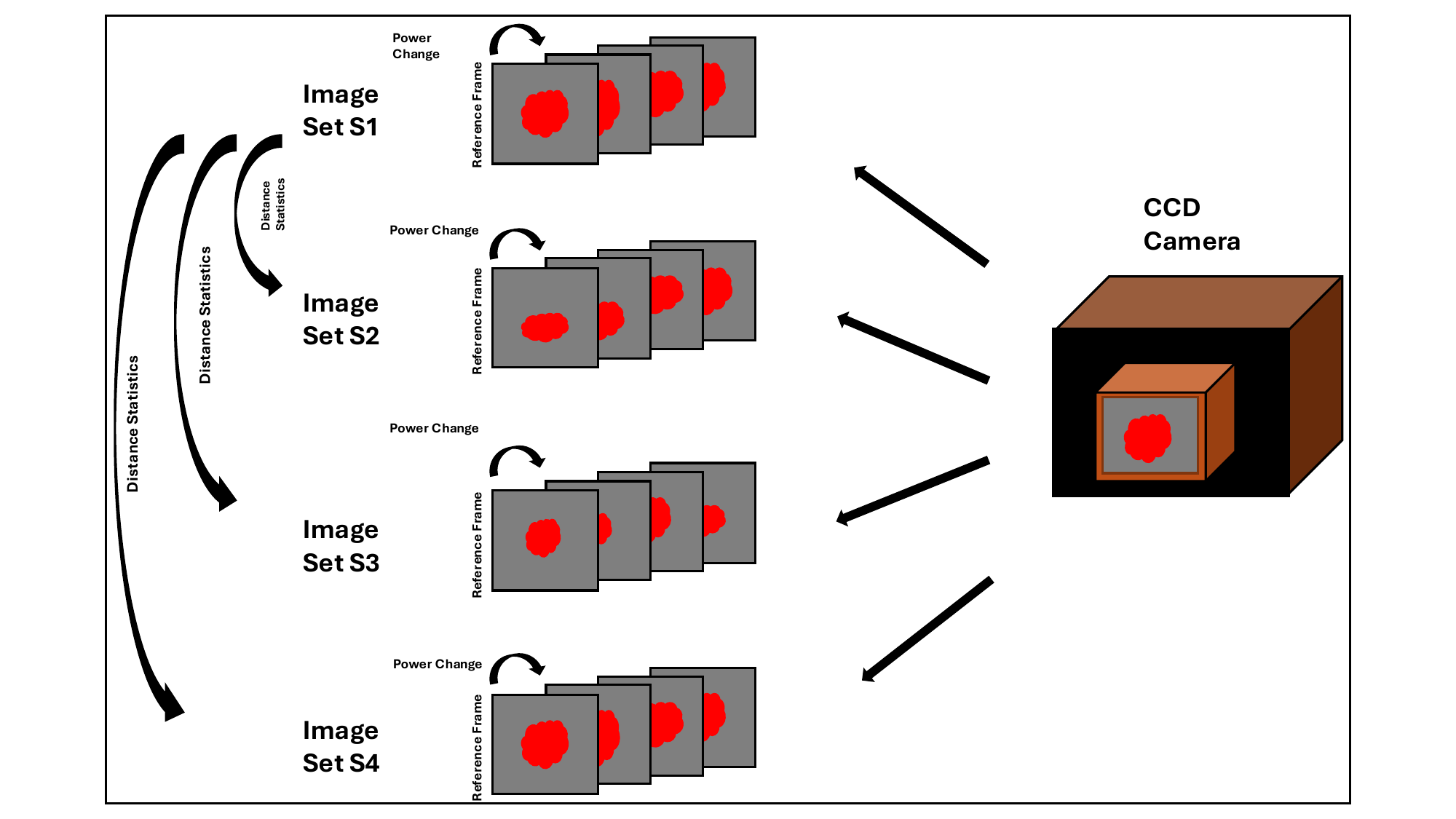}
    \subcaption{Block diagram Data Analysis Scheme II}
    \label{P1b}
\end{minipage}
\caption{Block diagrams illustrating the data analysis schemes.}
\label{P1}
\end{figure}

The data acquisition scheme (Fig.\ref{P1a} \& \ref{P1b}) captured beam fluctuations under four conditions: (i) baseline without turbulence, (ii) turbulence via PRPP only, (iii) turbulence with one PMMA rod, and (iv) turbulence with two rods. For each, 200 frames were recorded and analyzed by Gaussian fitting, with skewness and kurtosis (via Gram–Charlier expansion) quantifying deviations. This enabled direct statistical comparison of turbulence-only and turbulence–PMMA coupled cases.

Classical turbulence arises from irregular velocity fluctuations in viscous fluids such as the atmosphere, where flow can exist in either laminar or turbulent states. Laminar flow is smooth and orderly, while turbulence is dominated by random subflows, or eddies, that enhance mixing. The transition between these regimes is governed by the Reynolds number, $Re = Vl/\nu$, where $V$ is velocity, $l$ the flow scale, and $\nu$ the kinematic viscosity. When $Re$ exceeds a critical value (typically $\sim 10^5$ near the ground), turbulence develops. Kolmogorov’s theory describes turbulence as statistically homogeneous and isotropic at small scales, with large-scale energy generated by shear or convection cascading down to smaller eddies. Energy transfer occurs across an inertial range bounded by an outer scale $L_0$ and an inner scale $l_0$, until dissipation converts residual energy into heat. The corresponding turbulence power spectrum is expressed as
\begin{equation}
\Phi(k)=0.023r_0^{-5/3}k^{-11/3}.
\end{equation}
The Pseudo Random Phase Plate (PRPP) used in our experiment is a five-layered optical device designed to replicate atmospheric turbulence. It consists of two BK7 glass windows enclosing a central acrylic layer imprinted with a Kolmogorov-type turbulence profile. Near-index-matching polymer layers on each side of the acrylic provide mechanical stability, while the glass sealing enhances durability and minimizes environmental effects. The plate, about 10 mm thick, is robust and easily mounted on a rotary stage. It generates aberrated wavefronts with adjustable Fried coherence lengths ($r_0$ = 16–32 samples) across 4096 phase points, enabling controlled turbulence simulations Ref.\cite{62,91}.\par
\begin{figure}[H]
\centering
\begin{minipage}[b]{0.75\textwidth}
    \includegraphics[width=\textwidth]{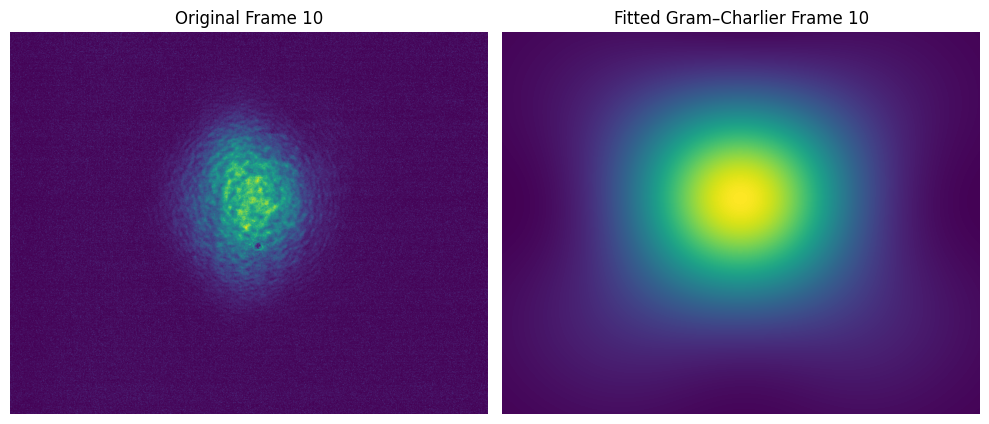}
    \caption{Original Image and Corresponding Fitted Image for Turbulence-free Beam}
    \label{R1a}
\end{minipage}
\end{figure}
\begin{figure}[H]
\centering
\begin{minipage}[b]{0.75\textwidth}
    \includegraphics[width=\textwidth]{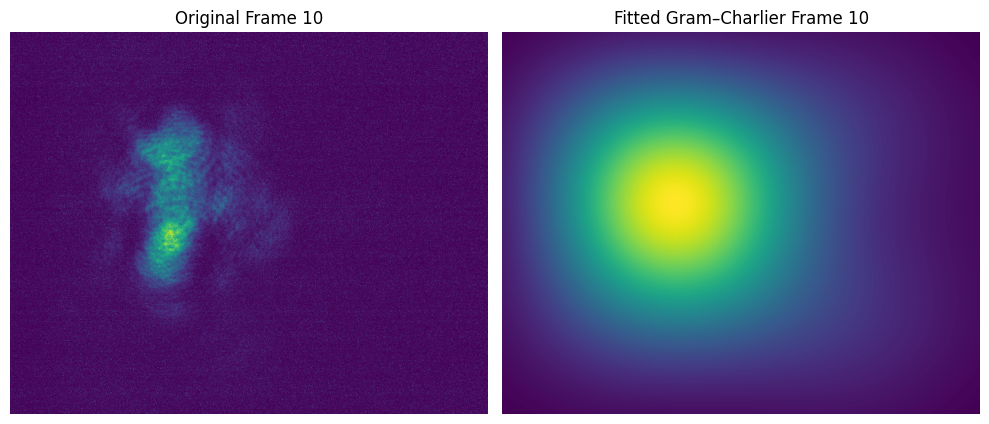}
    \caption{Original Image and Corresponding Fitted Image for Set 1: Raw Turbulence}
    \label{R1b}
\end{minipage}
\end{figure}
\begin{figure}[H]
\centering
\begin{minipage}[b]{0.75\textwidth}
    \includegraphics[width=\textwidth]{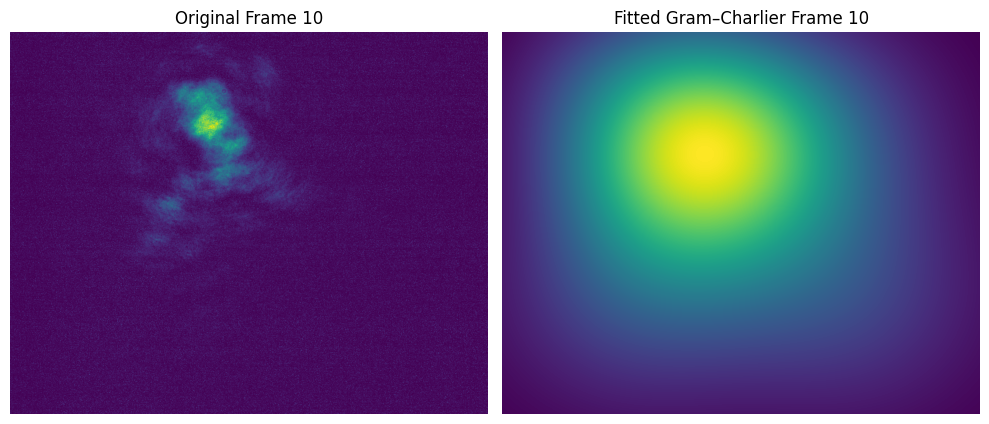}
    \caption{Original Image and Corresponding Fitted Image for Set 2: Turbulence with 1 PMMA Rod}
    \label{R1c}
\end{minipage}
\end{figure}
\begin{figure}[H]
\centering
\begin{minipage}[b]{0.75\textwidth}
    \includegraphics[width=\textwidth]{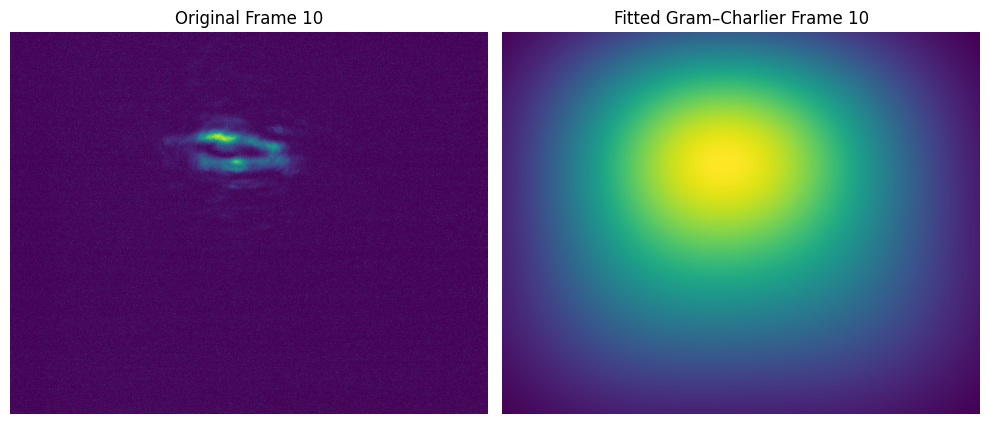}
    \caption{Original Image and Corresponding Fitted Image for Set 3: Turbulence with 2 PMMA Rod}
    \label{R1d}
\end{minipage}
\end{figure}

The recorded beam images were analyzed by fitting mathematical models to the experimental intensity distributions. Each figure presents the raw experimental image on the left, showing the beam’s spatial cross-section under a given condition, and the corresponding fitted image on the right, generated from a two-dimensional Gaussian or its Gram–Charlier expansion. The latter incorporates higher-order corrections through skewness (third-order cumulants) and kurtosis excess (fourth-order cumulants), enabling accurate representation of turbulence-induced deviations. In the turbulence-free case (Figure~\ref{R1a}), the beam exhibits a clean Gaussian profile with minimal distortion. Under raw turbulence without PMMA rods (Figure~\ref{R1b}), the images show strong asymmetry and spreading, reflected in higher skewness and kurtosis. With a single PMMA rod (Figure~\ref{R1c}), partial compensation occurs, reducing skewness though some excess kurtosis persists. With two rods (Figure~\ref{R1d}), turbulence effects are strongly suppressed, and the fitted profiles closely approach Gaussian behavior, confirming PMMA’s compensating role.



\section{Results Analysis and Discussion}\label{5}
In this section, we discuss the detailed results obtained from the statistical and topological analysis of turbulence-impacted beam propagation with and without PMMA slabs. The evaluation focuses on relative power fluctuations, statistical distributions, and correlation measures. The outcomes demonstrate how dipole-dipole coupling and gradient synchronization forces contribute to scintillation compensation and stability in beam propagation.\par

\begin{figure}[H]
\centering
\begin{minipage}[b]{0.75\textwidth}
    \includegraphics[width=\textwidth]{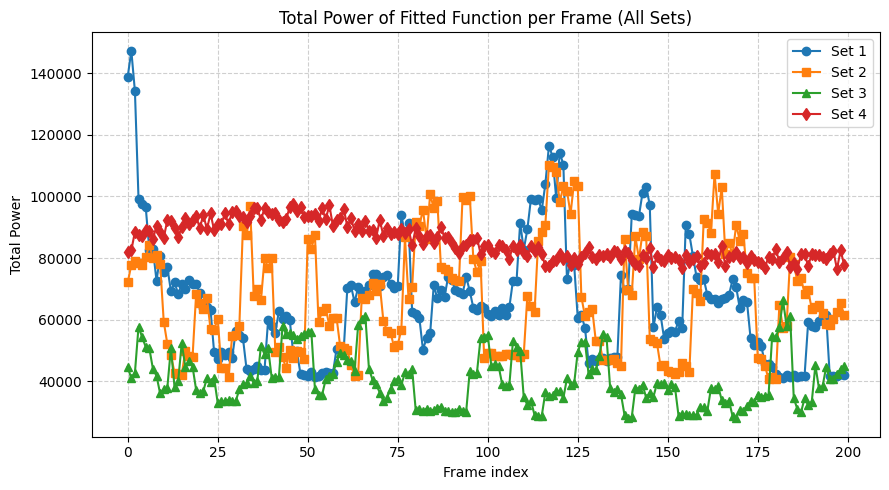}
    \caption{Power under Distorted Gaussian Fitted Functions. Set 1: Raw Turbulence, Set 2: Turbulence with 1 PMMA Rod, Set 3: Turbulence with 2 PMMA Rod and Set 4: Turbulence Free.}
    \label{R1}
\end{minipage}
\end{figure}

Figure~3 illustrates the total beam power computed under Gram-Charlier expanded Gaussian fittings for four conditions: turbulence-free, raw turbulence, turbulence with one PMMA rod, and turbulence with two PMMA rods. The turbulence-free case provides a baseline with minimal power variations across frames. In contrast, raw turbulence introduces significant irregular fluctuations, representing strong scintillation effects. The introduction of a single PMMA rod already reduces this fluctuation amplitude, while two rods further stabilize the power, highlighting enhanced compensation due to longer effective medium length. This confirms that coupled dipole oscillation synchronization suppresses higher-order distortions and restores stability to the power distribution.\par
Figure~4 shows relative power fluctuations computed with respect to the reference frame (0th frame). For raw turbulence, fluctuations are strong and widely spread, indicating loss of statistical consistency between consecutive frames. The presence of one PMMA rod reduces the range of fluctuations, but the variance is still notable. With two PMMA rods, the relative changes compress significantly, showing that the beam statistics remain more closely aligned with the reference condition. This emphasizes that medium-induced gradient forces damp turbulence-driven randomness and enforce collective mode stabilization.\par

\begin{figure}[H]
\centering
\begin{minipage}[b]{0.75\textwidth}
    \includegraphics[width=\textwidth]{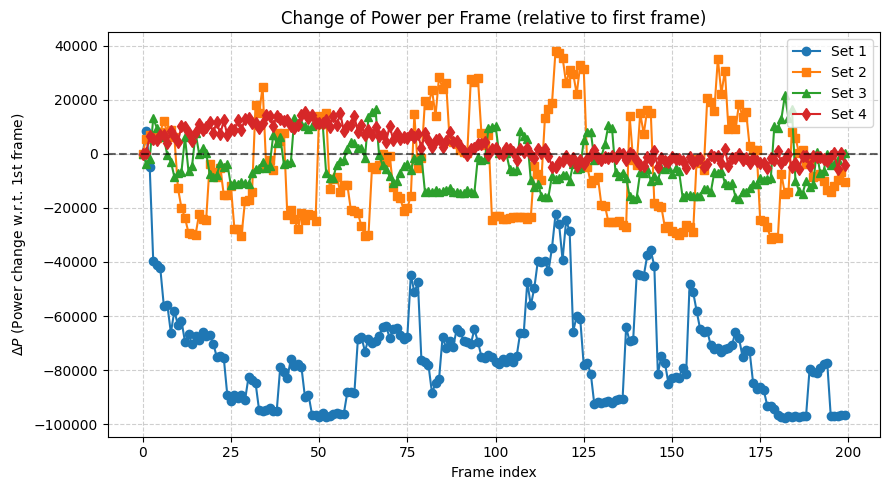}
    \caption{Relative Power Changes under Distorted Gaussian Fitted Functions. Set 1: Raw Turbulence, Set 2: Turbulence with 1 PMMA Rod, Set 3: Turbulence with 2 PMMA Rod and Set 4: Turbulence Free. Each pairs contains a reference frame (0th Frame) and a target frame (all other frames).}
    \label{R2}
\end{minipage}
\end{figure}
\begin{figure}[H]
\centering
\begin{minipage}[b]{0.75\textwidth}
    \includegraphics[width=\textwidth]{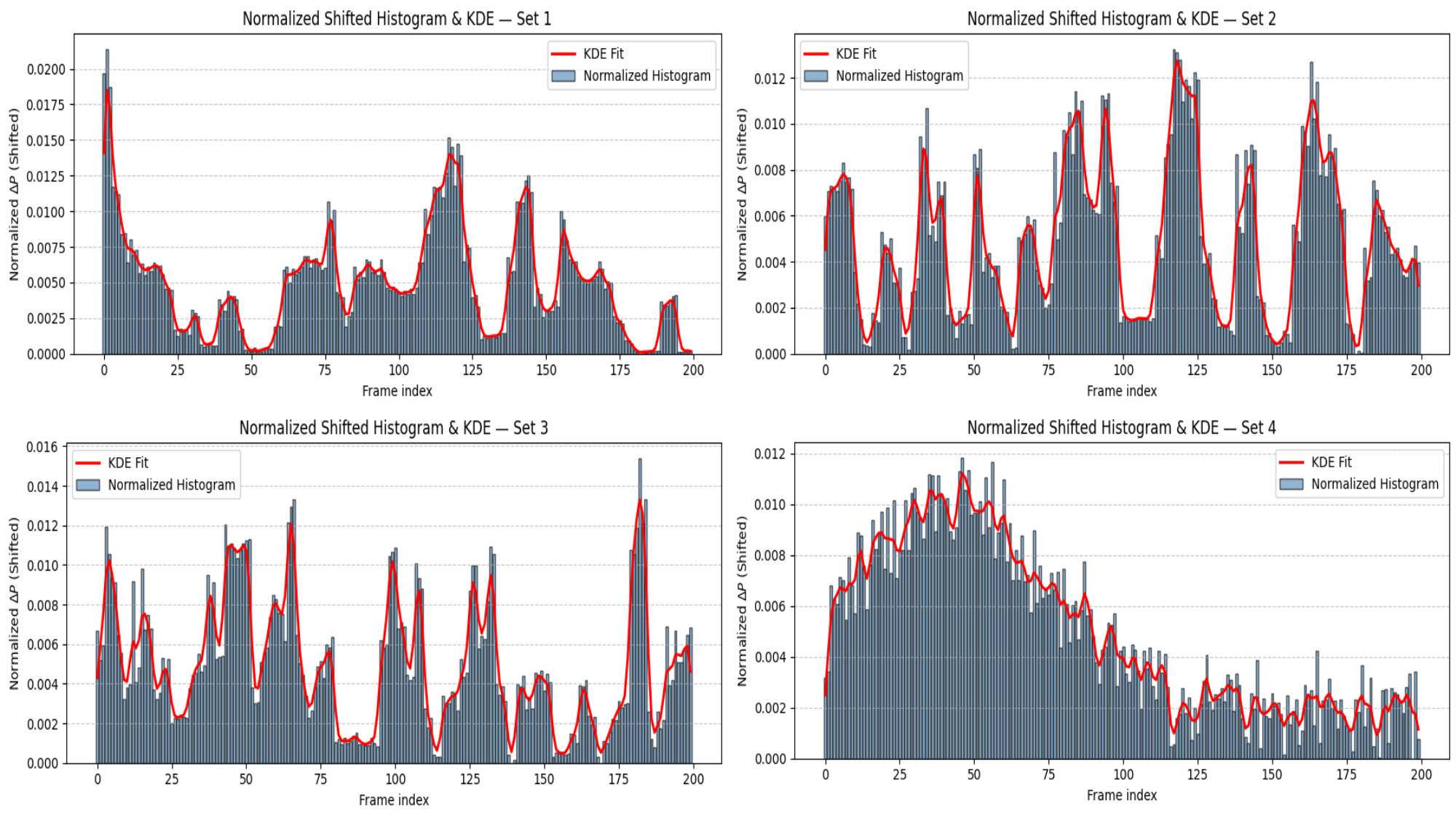}
    \caption{KDE Fitted Bar Diagram for Relative Power Fluctuations. Set 1: Raw Turbulence, Set 2: Turbulence with 1 PMMA Rod, Set 3: Turbulence with 2 PMMA Rod and Set 4: Turbulence Free. Each pairs contains a reference frame (0th Frame) and a target frame (all other frames).}
    \label{R3d}
\end{minipage}
\end{figure}
\begin{figure}[H]
\centering
\begin{minipage}[b]{0.75\textwidth}
    \includegraphics[width=\textwidth]{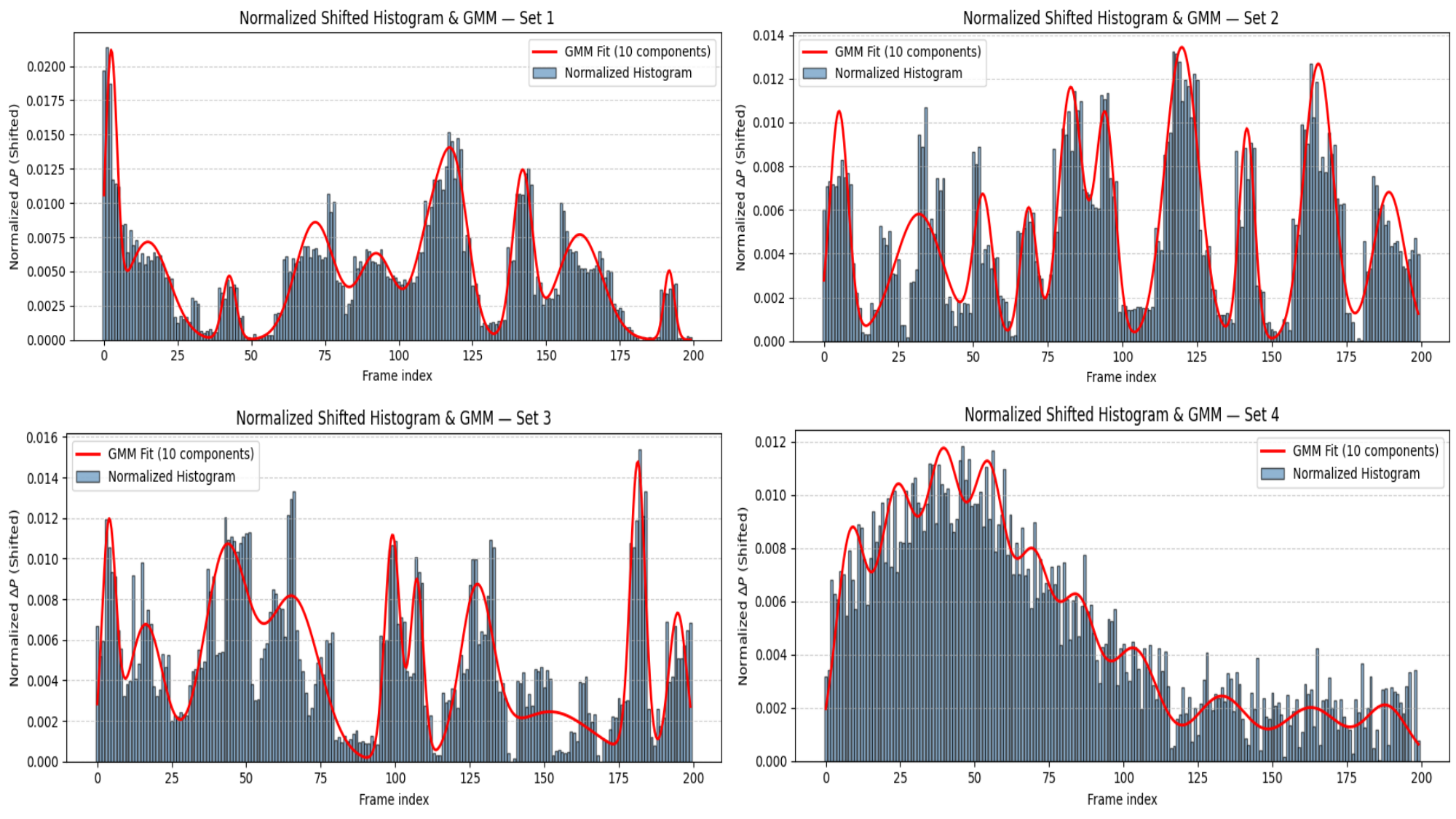}
    \caption{GMM Fitted Bar Diagram for Relative Power Fluctuations. Set 1: Raw Turbulence, Set 2: Turbulence with 1 PMMA Rod, Set 3: Turbulence with 2 PMMA Rod and Set 4: Turbulence Free. Each pairs contains a reference frame (0th Frame) and a target frame (all other frames).}
    \label{R3e}
\end{minipage}
\end{figure}

Figure~5 presents the normalized histograms of relative power fluctuations using Kernel Density Estimation (KDE). The turbulence-free distribution is narrow and unimodal, while raw turbulence exhibits broad, skewed distributions with heavy tails. With one PMMA rod, the KDE smooths out the distribution, though spread remains evident. Two PMMA rods sharpen the distribution, moving it closer to the turbulence-free condition. The KDE analysis demonstrates that additional medium length suppresses higher-order variability and recovers Gaussian-like stability in statistical behavior.\par
In Figure~6, Gaussian Mixture Model (GMM) fitting captures multimodal nature of the distributions. For raw turbulence, the fitting reveals multiple peaks, corresponding to unstable and fragmented beam statistics. The single PMMA rod case shows partial consolidation, with some components overlapping, while the two-rod case produces a dominant Gaussian component with reduced side modes. This confirms that turbulence compensation aligns the power fluctuations more closely to a unimodal Gaussian regime, reducing fragmentation of beam statistics.\par

\begin{figure}[H]
\centering
\begin{minipage}[b]{0.75\textwidth}
    \includegraphics[width=\textwidth]{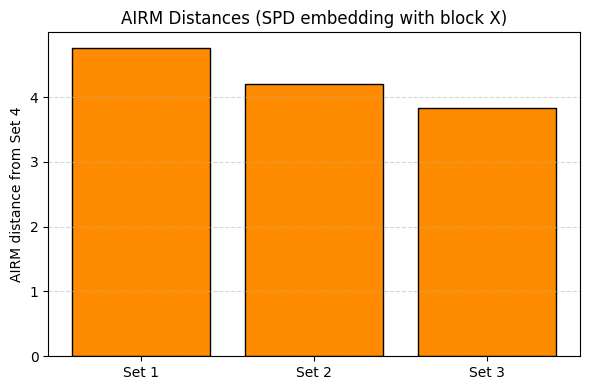}
    \caption{AIRM Topological Distance}
    \label{R4}
\end{minipage}
\end{figure}
\begin{figure}[H]
\centering
\begin{minipage}[b]{0.75\textwidth}
    \includegraphics[width=\textwidth]{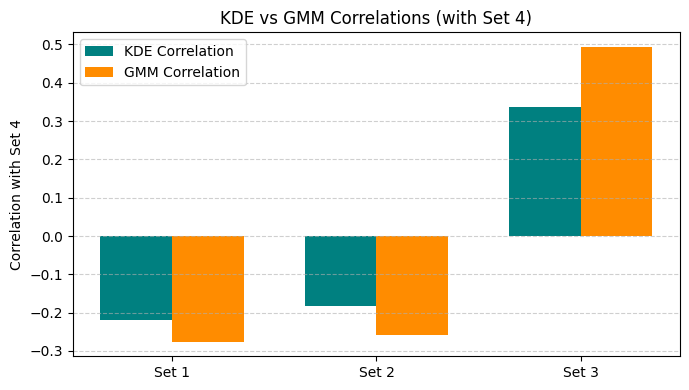}
    \caption{Pearson Correlation Coefficients}
    \label{R5}
\end{minipage}
\end{figure}

Figure~7 reports the statistical distances computed under the Affine-Invariant Riemannian Metric (AIRM). The raw turbulence condition yields the largest topological distances relative to the turbulence-free beam, indicating strong dissimilarity. With one PMMA rod, distances decrease, reflecting partial convergence. The two-rod case shows further contraction of distances, demonstrating that statistical topology of the beam with PMMA compensation becomes significantly closer to the turbulence-free case. This metric provides a rigorous quantification of how medium-induced dipole synchronization restores statistical similarity.\par
Finally, Figure~8 highlights Pearson correlation coefficients between turbulence-free and turbulence-impacted scenarios. For raw turbulence, the correlations are strongly negative, showing that fluctuations evolve in the opposite direction to the turbulence-free case. The introduction of one PMMA rod reduces the negativity, suggesting partial recovery of similarity. With two PMMA rods, the correlation turns positive, proving that beam fluctuations regain coherence with the turbulence-free reference. This result directly establishes that dipole coupling and gradient forces are responsible for synchronizing the dynamics and compensating turbulence-induced noise.

\section{Conclusion}\label{6}
This work has presented a unified theoretical and experimental framework for analyzing turbulence-impacted optical beam propagation using statistical topology and Lorentz dipole oscillation dynamics. By modeling beam intensity distributions through Gaussian Mixture Models and embedding them into the manifold of Symmetric Positive Definite matrices, we established a principled foundation for quantifying similarity, dissimilarity, and structural distortions induced by turbulence. The incorporation of affine-invariant Riemannian metrics and Pearson correlations enabled rigorous comparison across turbulence-free, raw turbulence, and PMMA-compensated scenarios. Experimental validation with one and two PMMA rods confirmed that coupled dipole oscillations and gradient synchronization forces significantly suppress scintillation, stabilize beam statistics, and restore statistical similarity to turbulence-free conditions. The KDE and GMM analyses demonstrated how compensation strategies reduce higher-order fluctuations and recover Gaussian-like stability, while distance and correlation measures provided quantitative evidence of restored coherence. Beyond its immediate relevance to adaptive and applied optics, the proposed framework opens avenues for secure free-space optical communication, turbulence-resilient imaging, and data compression techniques. Future efforts will extend the statistical geometry with refined GMM approaches, scatter-based metrics, and integration with machine learning methods for turbulence prediction and control. Overall, this study bridges theoretical modeling, experimental results, and technological applications, offering a versatile pathway to enhance the robustness of optical systems in random media.

\section*{Funding}
Department of Science and Technology, Ministry of Science and Technology, India (CRG/2020/003338).

\section*{Declaration of competing interest}
The authors declare the following financial interests/personal relationships which may be considered as potential competing interests: Shouvik Sadhukhan reports a relationship with Indian Institute of Space Science and Technology that includes: employment. NA If there are other authors, they declare that they have no known competing financial interests or personal relationships that could have appeared to influence the work reported in this paper.

\section*{Data availability}
All data used for this research has been provided in the manuscript itself.

\section*{Acknowledgments}
Shouvik Sadhukhan and C S Narayanamurthy Acknowledge the SERB/DST (Govt. Of India) for providing financial support via the project grant CRG/2020/003338 to carry out this work. Shouvik Sadhukhan would like to thank Mr. Amit Vishwakarma and Dr. Subrahamanian K S Moosath from Department of Mathematics, Indian Institute of Space Science and Technology Thiruvananthapuram for their suggestions into statistical analysis in this paper.

\section*{CRediT authorship contribution statement}
\textbf{Shouvik Sadhukhan:} Writing– original draft, Visualization, Formal analysis. \textbf{C. S. Narayanamurthy:} Writing– review $\&$ editing, Validation, Supervision, Resources, Project administration, Investigation, Funding acquisition, Conceptualization.


\end{document}